\documentclass[aps,pra,nobalancelastpage,twocolumn,superscriptaddress]{revtex4-1}

\usepackage{color,ifthen,amsthm,amsmath,amsxtra,amsfonts,dsfont,graphicx,bm,tikz,scalerel,wasysym, physics, bbm,  graphicx}
\usepackage[colorlinks=true,linkcolor=blue, citecolor=blue, urlcolor=blue, bookmarks]{hyperref}

\usepackage{amsthm}

\newcommand{\be}{\begin{equation}}
\newcommand{\ee}{\end{equation}}
\newcommand{\bea}{\begin{eqnarray}}
\newcommand{\eea}{\end{eqnarray}}
\newcommand{\bse}{\begin{subequations}}
\newcommand{\ese}{\end{subequations}}

\theoremstyle{plain}
\newcommand{\1}{\mathbbm{1}}
\newcommand{\bondD}{\mathcal{D}}

\usetikzlibrary{arrows,calc,matrix,backgrounds,fit,positioning,decorations.pathmorphing,decorations.markings,patterns,decorations.pathreplacing,shapes.misc}
\definecolor{FcolU}{rgb}{0.71,0.78,0.91}
\definecolor{colMPS}{rgb}{0.27,0.45,0.77}
\definecolor{colVMPS}{rgb}{0.96,0.74,0.59}
\definecolor{colLines}{rgb}{0.31,0.31,0.31}
\definecolor{colMPSLines}{rgb}{0,0.01,0.18}
\definecolor{colVMPSLines}{rgb}{0.11,0.11,0.11}
\definecolor{IcolUc}{rgb}{0.71,0.41,0.42}
\definecolor{IcolU}{rgb}{0.71,0.8,0.76}
\definecolor{IcolVMPSc}{rgb}{0.73,0.69,0.7}
\definecolor{IcolVMPS}{rgb}{0.81,0.77,0.78}
\definecolor{colObs}{rgb}{1.,1.,1.}

\def\dx{0.3}
\def\r{0.08}
\def\sqrtThree{1.7320508075688772}

\newcommand\mpsWire[4]{
	\draw [very thick,colMPSLines] ({(#2)*\dx},{-(#1)*\dx}) -- ({(#4)*\dx},{-(#3)*\dx});
}

\newcommand\gridLine[4]{
  \draw [colLines] ({(#2)*\dx},{-(#1)*\dx}) -- ({(#4)*\dx},{-(#3)*\dx});
}

\newcommand\rcaGate[4]{
	\draw [thick,rounded corners=0.5,colLines,fill=#4] 
	({\dx*#2-0.5*\r},{-\dx*#1+0.25*\dx}) rectangle ({\dx*#3+0.5*\r},{-\dx*#1-0.25*\dx});
}

\newcommand\IvmpsWire[4]{
	\draw [double,colVMPSLines]({(#2)*\dx},{-(#1)*\dx}) -- ({(#4)*\dx},{-(#3)*\dx});
}

\newcommand\vmpsWire[4]{
	\draw [double,colVMPSLines]({(#2)*\dx},{-(#1)*\dx}) -- ({(#4)*\dx},{-(#3)*\dx});
}

\newcommand\vmpsV[3]{
	\draw [thick,rounded corners=0.5,colVMPSLines,fill=#3] ({\dx*#2},{-\dx*#1})
    +({\r/2.},{-\sqrtThree*\r/2.}) -- +({\r/2.},{\sqrtThree*\r/2.}) 
	-- +({-\r},0) -- cycle;
}

\newcommand\vmpsW[3]{
	\draw [thick,rounded corners=0.5,colVMPSLines,fill=#3] ({\dx*#2},{-\dx*#1})
    +({-\r/2.},{-\sqrtThree*\r/2.}) -- +({-\r/2.},{\sqrtThree*\r/2.}) 
    -- +({\r},0) -- cycle;
}

\newcommand\mpsBvecV[3]{
	\draw [thick, rounded corners=0.5,colMPSLines,fill=#3] ({\dx*#2},{-\dx*#1})
    +({\r/2.},{-\sqrtThree*\r/2.}) -- +({\r/2.},{\sqrtThree*\r/2.}) 
	-- +({-\r},0) -- cycle;
	\node at ({\dx*#2},{-\dx*#1}) {$\cdot$};
}

\newcommand\mpsBvecW[3]{
	\draw [thick, rounded corners=0.5,colMPSLines,fill=#3] ({\dx*#2},{-\dx*#1})
    +({-\r/2.},{-\sqrtThree*\r/2.}) -- +({-\r/2.},{\sqrtThree*\r/2.}) 
	-- +({\r},0) -- cycle;
	\node at ({\dx*#2},{-\dx*#1}) {$\cdot$};
}

\newcommand\mpsBvec[2]{
	\draw [very thick,colMPSLines] ({\dx*#2},{-\dx*#1}) + ({-0.75*\r},0) -- + ({0.75*\r},0);
}

\newcommand\mpsA[3]{
	\draw [thick,rounded corners=0.5,colMPSLines,fill=#3,rotate around={45:({\dx*#2},{-\dx*#1})}] 
	({\dx*#2},{-\dx*#1}) +({0.75*\r},{0.75*\r}) rectangle +({-0.75*\r},{-0.75*\r});
}

\newcommand\mpsB[3]{
	\draw [thick,rounded corners=0.5,colMPSLines,fill=#3] ({\dx*#2},{-\dx*#1}) +({0.75*\r},{0.75*\r}) rectangle +({-0.75*\r},{-0.75*\r});
}

\newcommand\mpsC[4]{
	\draw [thick,rounded corners=0.5,colMPSLines,fill=#4] 
	({\dx*#3+0.75*\r},{-\dx*#1+0.75*\r}) rectangle ({\dx*#3-0.75*\r},{-\dx*#2-0.75*\r});
	\draw [thick] ({\dx*#3-0.75*\r},{-\dx*(#1+#2)/2}) -- ({\dx*#3+0.75*\r},{-\dx*(#1+#2)/2});
}

\newcommand\bCircle[3]{
	\draw [thick,colLines,fill=#3] ({\dx*#2},{-\dx*#1}) circle ({1.25*\r});
}

\newcommand\sCircle[3]{
	\draw [thick,colLines,fill=#3] ({\dx*#2},{-\dx*#1}) circle ({0.5*\r});
}

\newcommand\extSquare[4]{
	\draw [thick,rounded corners=0.5,colLines,fill=#4] ({\dx*#2-1.25*\r},{-\dx*#1-1.25*\r}) rectangle ({\dx*#3+1.25*\r},{-\dx*#1+1.25*\r});
}

\newcommand\nclME[2]{
	\draw [thick,colLines,fill=white] ({#2*\dx},{-#1*\dx}) circle ({0.5*\r});
}

\newcommand\nME[2]{
	\draw [thick,fill=gray] ({#2*\dx},{-#1*\dx}) circle ({0.5*\r});
}

\newcommand\ngridLine[4]{
	\draw [thick,colLines] ({(#2)*\dx},{-(#1)*\dx}) -- ({(#4)*\dx},{-(#3)*\dx});
}

\newcommand\leftHook[2]{
	\draw ({#2*\dx},{-\dx*#1}) arc (90:240:{0.125*\dx});
}

\newcommand\rightHook[2]{
	\draw ({#2*\dx},{-\dx*#1}) arc (90:-50:{0.125*\dx});
}

\newcommand\TEsheet[2]{
  \draw [thick,colLines,fill=IcolU,rounded corners=0.5] ({(#2)*\dx},{-(#1)*\dx}) rectangle ({((#2)+7)*\dx},{(-(#1)+4)*\dx});
  \draw [thick,colVMPSLines,fill=IcolVMPS,rounded corners=0.5] ({((#2)-0.01)*\dx},{(-(#1)-0.01)*\dx}) rectangle ({((#2)+7+0.01)*\dx},{(-(#1)+0.31)*\dx});
}

\newcommand\TECsheet[2]{
  \draw [thick,colLines,fill=IcolUc,rounded corners=0.5] ({(#2)*\dx},{-(#1)*\dx}) rectangle ({((#2)+7)*\dx},{(-(#1)+4)*\dx});
  \draw [thick,colVMPSLines,fill=IcolVMPSc,rounded corners=0.5] ({((#2)-0.01)*\dx},{(-(#1)-0.01)*\dx}) rectangle ({((#2)+7+0.01)*\dx},{(-(#1)+0.31)*\dx});
}

\newcommand\connection[4]{
  \draw [thick,colLines,fill=white,rounded corners=0.5] ({((#4)-0.01+3.5)*\dx},{-((#3)+0.01-4)*\dx}) rectangle ({((#4)+0.01)*\dx},{-((#3)-0.25-4)*\dx});
  \draw [thick,colLines,fill=white,rounded corners=0.5] ({((#4)-0.01+3.5)*\dx},{-((#3)-0.25-4)*\dx}) -- ({((#2)-0.01+3.5)*\dx},{-((#1)-0.25-4)*\dx}) -- ({((#2)+0.01)*\dx},{-((#1)-0.25-4)*\dx}) -- ({((#4)+0.01)*\dx},{-((#3)-0.25-4)*\dx}) -- cycle;
  \draw [thick,colLines,fill=white,rounded corners=0.5] 
  ({((#2)-0.01+3.5)*\dx},{-((#1)-0.25-4)*\dx}) --
  ({((#2)+0.01)*\dx},{-((#1)-0.25-4)*\dx}) -- 
  ({((#2)+0.01)*\dx},{-((#1)-0.01-4)*\dx}) -- 
  ({((#2)-0.01+3.5)*\dx},{-((#1)-0.01-4)*\dx}) -- cycle;
}

\newcommand\bConnection[4]{
  \draw [thick,colLines,fill=white,rounded corners=0.5] ({((#4)-0.01+3.5)*\dx},{-((#3)-0.25-4)*\dx}) -- ({((#2)-0.01+3.5)*\dx},{-((#1)-0.25-4)*\dx}) -- ({((#2)+0.01)*\dx},{-((#1)-0.25-4)*\dx}) -- ({((#4)+0.01)*\dx},{-((#3)-0.25-4)*\dx}) -- cycle;
  \draw [thick,colLines,fill=white,rounded corners=0.5] 
  ({((#2)-0.01+3.5)*\dx},{-((#1)-0.25-4)*\dx}) --
  ({((#2)+0.01)*\dx},{-((#1)-0.25-4)*\dx}) -- 
  ({((#2)+0.01)*\dx},{-((#1)-0.01-4)*\dx}) -- 
  ({((#2)-0.01+3.5)*\dx},{-((#1)-0.01-4)*\dx}) -- cycle;
}

\newcommand\fConnection[4]{
  \draw [thick,colLines,fill=white,rounded corners=0.5] ({((#4)-0.01+3.5)*\dx},{-((#3)+0.01-4)*\dx}) rectangle ({((#4)+0.01)*\dx},{-((#3)-0.25-4)*\dx});
  \draw [thick,colLines,fill=white,rounded corners=0.5] ({((#4)-0.01+3.5)*\dx},{-((#3)-0.25-4)*\dx}) -- ({((#2)-0.01+3.5)*\dx},{-((#1)-0.25-4)*\dx}) -- ({((#2)+0.01)*\dx},{-((#1)-0.25-4)*\dx}) -- ({((#4)+0.01)*\dx},{-((#3)-0.25-4)*\dx}) -- cycle;
}

\theoremstyle{plain}

\begin{document}

\title{Exact Thermalization Dynamics in the ``Rule $54$'' Quantum Cellular Automaton}

\author{Katja Klobas}
\affiliation{Rudolf Peierls Centre for Theoretical Physics, Clarendon Laboratory, Oxford University, Parks Road, Oxford OX1 3PU, United Kingdom}

\author{Bruno Bertini}
\affiliation{Rudolf Peierls Centre for Theoretical Physics, Clarendon Laboratory, Oxford University, Parks Road, Oxford OX1 3PU, United Kingdom}

\author{Lorenzo Piroli}
\affiliation{Max-Planck-Institut f\"ur Quantenoptik, Hans-Kopfermann-Str. 1, 85748 Garching, Germany}
\affiliation{Munich Center for Quantum Science and Technology, Schellingstra\ss e 4, 80799 M\"unchen, Germany}

\begin{abstract}
We study the out-of-equilibrium dynamics of the quantum cellular automaton known as ``Rule 54". For a class of low-entangled initial states, we provide an analytic description of the effect of the global evolution on finite subsystems in terms of simple quantum channels, which gives access to the full thermalization dynamics at the microscopic level. As an example, we provide analytic formulae for the evolution of local observables and R\'enyi entropies. We show that, in contrast to other known examples of exactly solvable quantum circuits, Rule 54 does not behave as a simple Markovian bath on its own parts, and displays typical non-equilibrium features of interacting integrable many-body quantum systems such as finite relaxation rate and interaction-induced dressing effects. Our study provides a rare example where the full thermalization dynamics can be solved exactly at the microscopic level. 
\end{abstract}

\maketitle

When a generic isolated quantum many-body system is driven out of equilibrium, its local properties are eventually described by the thermal ensemble. This picture can be intuitively explained by saying that, in the thermodynamic limit, the system acts as a bath for its own local sub-systems~\cite{rigol2008thermalization,cazalilla2010focus,Polkovnikov2011,d2016quantum,essler2016quench}. In light of the undeniable success of this paradigm, it is perhaps surprising that for interacting systems most of the evidence in its support comes from numerical computations in relatively small systems. The reason is that computing the full many-body relaxation dynamics in the presence of interactions poses formidable challenges that are difficult to overcome even in ``exactly solvable'' systems like quantum integrable models~\cite{calabrese2016introduction,essler2016quench,caux2016quench}.

\begin{figure}
 \includegraphics[width=\columnwidth]{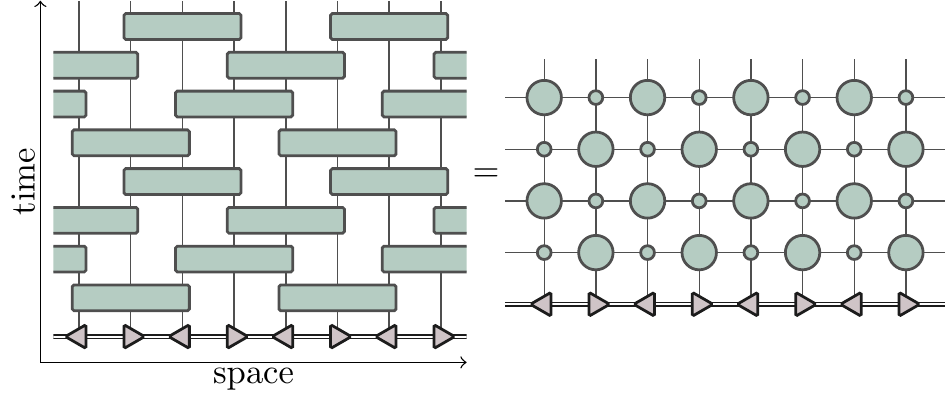}
\caption{Quantum circuit representation of the Rule 54 QCA. The dynamics can be
equivalently given in terms of three-site gates (left) and as an MPO (right).}
\label{fig:circuit}
\end{figure}

Recently, a useful arena to construct tractable models of many-body physics out of equilibrium has been identified in \emph{quantum circuits}. In the simplest setting, one considers a one-dimensional ($1D$) set of qudits evolved by a ``brickwork'' circuit, where a given initial state is updated by sequences of  local unitary gates (cf. Fig.~\ref{fig:circuit}). While in general quantum circuits offer no simplification with respect to  local-Hamiltonian dynamics~\footnote{In fact, any continuous evolution driven by a local-Hamiltonian can be approximated arbitrarily well by a quantum circuit~\cite{osborne2006efficient}}, the discreteness of the evolution makes it  possible to construct non-trivial solvable models, with notable examples given by random~\cite{chan2018solution, nahum2017quantum,nahum2018operator,vonKeyserlingk2018operator} and dual unitary circuits~\cite{bertini2019exact}. These systems proved to be useful minimal models for the quantum chaotic dynamics, enabling the analysis of aspects that are notoriously hard to tackle in traditional systems~\cite{bertini2019entanglement, bertini2019exact,gopalakrishnan2019unitary,bertini2020operatorI,bertini2020operatorII,bertini2020scrambling,piroli2020exact,claeys2020maximum,gutkin2020local,rather2020creating, kos2020correlations, rakovszky2018diffusive, khemani2018operator, chan2018spectral, suenderhauf2018localization, friedman2019spectral, li2019measurement, skinner2019measurement, zabalo2020critical, bao2020theory, choi2020quantum, lavasani2021measurementinduced, kos2020chaos, claeys2020ergodic, chan2018solution, nahum2017quantum,nahum2018operator,vonKeyserlingk2018operator, rakovszky2019sub}. However, when seen as models for thermalization, they are not typical: for instance, in dual-unitary circuits the action of the global evolution on any subsystem is purely Markovian, even in the absence of noise~\cite{bertini2019entanglement, piroli2020exact,lerose2020influence}. 

In this Letter, we present an exact solution for the thermalization dynamics in a quantum circuit that provably exhibits typical features of interacting many-body systems: the quantum version of the ``Rule 54'' cellular automaton~\cite{bobenko1993two}. The latter can be regarded as the simplest interacting integrable system and over the last years has been shown to provide an ideal ground for studying interacting many-body dynamics ~\cite{prosen2016integrability,gopalakrishnan2018operator,gopalakrishnan2018hydrodynamics,klobas2019time,buca2019exact,alba2019operator, klobas2020space, klobas2020matrix, prosen2017exact, inoue2018two, alba2020diffusion, buca2019exact, Friedman2019Integrable, hillberry2020entangled, gopalakrishnan2018facilitated}.

Our approach is based on a general tensor network (TN) algorithm introduced in Ref.~\cite{banuls2009matrix,muller2012tensor} (see also \cite{hastings2015connecting}) to describe the evolution of any subsystem in the thermodynamic limit. Specifically, we identify a set of algebraic relations obeyed by the tensors of Rule 54 that enable us to follow such an algorithm analytically.
We use this to derive exact formulae for the evolution of local observables, two-point correlation functions and R\'enyi entropies, providing a rare example where the full microscopic dynamics can be solved exactly, beyond non-interacting models and the perfectly Markovian regime.

Rule $54$ is defined by a $1D$ lattice of qubits where the time evolution is discrete and generated by the unitary operator $\mathbb{U}=\mathbb{U}_{\rm e}\mathbb{U}_{\rm o}$, with $\mathbb{U}_{\rm e}=\prod_{j} U_{2j-1,2j,2j+1}$, $\mathbb{U}_{\rm o}=\prod_{j} U_{2j,2j+1,2j+2}$. In the local computational basis $\{\ket{s_j}_j\}_{s_j=0,1}$, the matrix elements of the three-site unitary gate $U_{j-1,j,j+1}$ read
\be
U_{s_{1}^{\phantom{\prime}} s_{2}^{\phantom{\prime}} s_{3}^{\phantom{\prime}}}^{s_{1}^{\prime} s_{2}^{\prime} s_{3}^{\prime}}=\delta_{s_{1}, s_{1}^{\prime}} \delta_{\chi\left(s_{1}, s_{2}, s_{3}\right), s_{2}^{\prime}} \delta_{s_{3}, s_{3}^{\prime}}\,,
\label{eq:gateU}
\ee
where $\chi\left(s_{1}, s_{2}, s_{3}\right) \equiv (s_{1}+s_{2}+s_{3}+s_{1} s_{3})\bmod 2$. Note that this update rule causes a non-zero scattering shift for quasiparticles~\cite{bobenko1993two}. The operators $U_{j-1,j,j+1}$ and $U_{j+1,j+2,j+3}$ commute, which allows us to write $\mathbb{U}$ in the form of a brickwork quantum circuit (Fig.~\ref{fig:circuit}). We consider periodic chains of length $2L$, eventually taking  $L\to\infty$. 

We study a quench protocol~\cite{calabrese2006time,calabrese2007quantum} where the system is initialised in a low-entangled state, which we take to be a matrix product state (MPS)~\cite{fannes1992finitely,perez2007matrix} ${\ket{\Psi_0}=\sum_{i_{1}, \ldots, i_{2L}=0,1} \operatorname{tr}\left(A_{1}^{i_{1}} \ldots A_{2L}^{i_{2L}}\right)\ket{i_{1}, \ldots, i_{2L}}}$, 
where $A_j$ are $\bondD$-dimensional matrices ($\bondD$ is called the bond dimension). 

It is straightforward to see that $\mathbb{U}_{\rm e}$, $\mathbb{U}_{\rm o}$ can be represented as two-site shift invariant matrix product operators (MPO) with ${\bondD=2}$~\footnote{See the Supplemental Material}, so that the evolution can be computed by applying a sequence of MPOs to $|\Psi(0)\rangle$. Note that this representation is completely general, since any quantum cellular automata can be expressed exactly as an MPO with finite bond-dimension~\cite{cirac2017matrix,sahinoglu2018matrix,piroli2020quantum}.

\begin{figure}
    \includegraphics[width=\columnwidth]{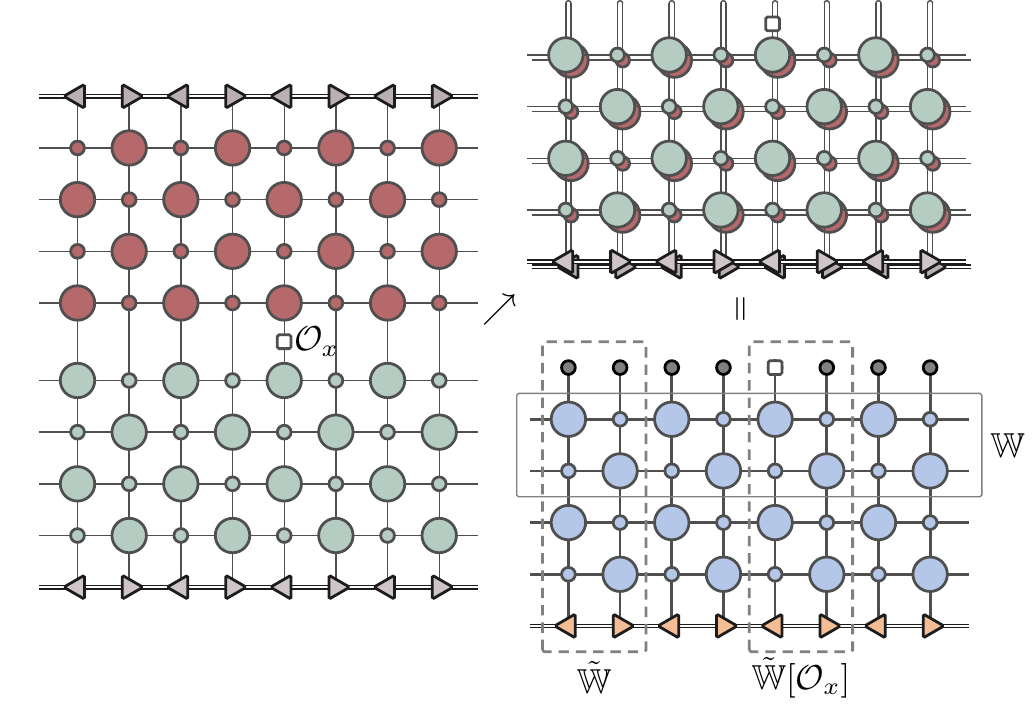}
  \caption{Operator evolution in the MPO form starting from an MPS and the folding procedure. Time-evolution of operators can be efficiently represented by combining the tensors and their complex conjugate into a \emph{super-tensor} acting on the doubled space. \label{fig:TN}}
\end{figure}

The expectation value of a local observable $\bra{\Psi(t)}\mathcal{O}_x\ket{\Psi(t)}$ evolving via MPOs is naturally represented by the TN depicted in Fig.~\ref{fig:TN}. In fact, it is convenient to think of such an object in the so-called \emph{folded} representation~\cite{banuls2009matrix}, where the original TN is bent in half so that each tensor associated with $U^\dagger$ ends up lying on top of the corresponding tensor of $U$. This procedure yields a new TN generated by a \emph{folded transfer matrix} ${\mathbb{W}}$, where the dimensions of local and auxiliary degrees of freedom are squared. These steps are depicted in Fig.~\ref{fig:TN}, where $\begin{tikzpicture}
    [baseline={([yshift=-0.6ex]current bounding box.center)},scale=1.75]
    \ngridLine{0}{0}{0.5}{0};
    \nME{0}{0};
  \end{tikzpicture}=
  \begin{bmatrix} 1 & 0 & 0 & 1 \end{bmatrix}$, represents the folded identity operator and the dynamics are defined by the folded tensors
\be 
\begin{aligned}
\begin{tikzpicture}
  [baseline={([yshift=-0.6ex]current bounding box.center)},scale=1.75]
  \ngridLine{0.75}{0}{-0.75}{0};
  \ngridLine{0}{0.75}{0}{-0.75};
  \bCircle{0}{0}{FcolU}
  \node at ({-1.25*\dx},{0}) {\scalebox{0.8}{$s_1 r_1$}};
  \node at (0,{-\dx}) {\scalebox{0.8}{$s_2 r_2$}};
  \node at ({1.25*\dx},{0}) {\scalebox{0.8}{$s_3 r_3$}};
  \node at (0,{\dx}) {\scalebox{0.8}{$s_4 r_4$}};
\end{tikzpicture}&=
  \delta_{\chi(s_1,s_2,s_3),s_4}
  \delta_{\chi(r_1,r_4,r_3),r_2}
  ,\\
\begin{tikzpicture}
  [baseline={([yshift=-0.6ex]current bounding box.center)},scale=1.75]
  \ngridLine{0.75}{0}{-0.75}{0};
  \ngridLine{0}{0.75}{0}{-0.75};
  \sCircle{0}{0}{FcolU}
  \node at ({-1.25*\dx},{0}) {\scalebox{0.8}{$s_1 r_1$}};
  \node at (0,{-\dx}) {\scalebox{0.8}{$s_2 r_2$}};
  \node at ({1.25*\dx},{0}) {\scalebox{0.8}{$s_3 r_3$}};
  \node at (0,{\dx}) {\scalebox{0.8}{$s_4 r_4$}};
\end{tikzpicture}&=\prod_{j=1}^3 \delta_{s_{j},s_{j+1}}
\delta_{r_{j},r_{j+1}}.
\end{aligned}
\ee
At this point, following Refs.~\cite{banuls2009matrix,muller2012tensor}, it is instructive to look at the expectation value in the \emph{t-channel}. Namely, to view the diagram in Fig.~\ref{fig:TN} as the TN formed by the product of $2L$ MPOs acting on the lattice in time and propagating in space (a similar t-channel description has found useful applications also in the study of spectral properties~\cite{bertini2018exact,flack2020statistics,braun2020transition,garratt2020many}). Specifically we have 
\be
\label{eq:transverse}
\bra{\Psi(t)}\mathcal{O}_x\ket{\Psi(t)}=\operatorname{tr}\left[\tilde{\mathbb{W}}^{L-1} \tilde{\mathbb{W}}[\mathcal{O}_x]\right]\,.
\ee
Here, we denoted by $\tilde{\mathbb{W}}$ the MPO encoding the evolution along the ``space direction'', while $\tilde{\mathbb{W}}[{\mathcal{O}}_x]$ corresponds to the transfer matrix associated with the application of the single-site operator $\mathcal{O}_x$, cf. Fig.~\ref{fig:TN}. It is straightforward to show that $\tilde{\mathbb{W}}$ has a unique largest eigenvalue $\bar \lambda=1$~\cite{Note2}. Exploiting this fact, we can evaluate Eq.~\eqref{eq:transverse} in the thermodynamic limit, obtaining
\be
\lim _{L \rightarrow \infty}\bra{\Psi(t)}\mathcal{O}_x\ket{\Psi(t)}=\langle L|\tilde{\mathbb{W}}[\mathcal{O}_x]| R\rangle\,,
\label{eq:td_contraction}
\ee
where $\bra{L}$ and $\ket{R}$ denote, respectively, the left and right fixed points of $\tilde{\mathbb{W}}$ (i.e. eigenvectors associated with $\bar \lambda$). The above reasoning can be repeated for local operators of any finite support and implies that the fixed points encode the action of the whole system on all finite subsystems, playing the role of an effective reservoir, and hence contain \emph{all} information about local relaxation. In general, however, $\bra{L}$ and $\ket{R}$ can only be obtained numerically, with a computational cost that increases exponentially with time~\cite{banuls2009matrix,muller2012tensor}. Here we show that, in Rule 54, $\bra{L}$ and $\ket{R}$ can be obtained \emph{analytically} for all times.

Before proceeding, it is important to note that $\bra{L}$ and $\ket{R}$ generally bear a strong dependence on the initial state. 
Our goal is to identify initial MPSs that thermalize, i.e., whose local properties at large times approach those of an infinite-temperature state. One therefore expects the fixed points of $\tilde{\mathbb{W}}$ to be similar to to those of the transfer matrix~$\tilde{\mathbb{W}}_\infty$, corresponding to the infinite temperature state [Note that the correlation functions on the infinite temperature state are given by the TN analogous to the one in Fig.~\ref{fig:TN}, with the initial state replaced by the (appropriately normalised) identity operator]. Thus, we begin by analysing the latter. 

As our first main result, we show that the fixed points of $\tilde{\mathbb{W}}_{\infty}$ can be computed analytically: this  allows us to compute efficiently \emph{all} infinite-temperature multi-point correlation functions supported in a finite interval, in a sense that will be made precise later [cf. Eq.~\eqref{eq:channelC} and subsequent discussions]. This result generalises the findings of Ref.~\cite{klobas2020space} to the quantum case. Specifically, we find that the leading left and right eigenvectors can be expressed  in terms of the $3\times 3$ local tensors $A_{s r}$, $B_{s r}$ and the boundary vector $\ket{b}$ (the explicit expressions are reported in the Supplemental Material)
\begin{equation}
\label{eq:boundaryvectors}
  \begin{tikzpicture}
    [baseline={([yshift=-0.6ex]current bounding box.center)},scale=1.75]
    \mpsWire{-0.5}{0}{.5}{0};
    \ngridLine{0}{0}{0}{0.5};
    \mpsA{0}{0}{colMPS}
    \node at ({0.875*\dx},0) {\scalebox{0.75}{$s\,r$}};
  \end{tikzpicture} = A_{sr},\quad
  \begin{tikzpicture}[baseline={([yshift=-0.6ex]current bounding box.center)},scale=1.75]
    \mpsWire{-0.5}{2}{.5}{2};
    \ngridLine{0}{2}{0}{2.5};
    \mpsB{0}{2}{colMPS};
    \node at ({2.875*\dx},0) {\scalebox{0.75}{$s\,r$}};
  \end{tikzpicture} = B_{sr},\quad
  \begin{tikzpicture}[baseline={([yshift=-0.6ex]current bounding box.center)},scale=1.75]
    \mpsBvec{0}{0};
    \mpsWire{0}{0}{-0.5}{0};
  \end{tikzpicture} = \ket{b},
\end{equation}
that fulfil the following set of local relations, which we term ``zipping conditions"
\begin{subequations}\label{eq:local_identities}
  \begin{align}
    \label{eq:MagicRelations1}
    \begin{tikzpicture}
      [baseline={([yshift=-0.6ex]current bounding box.center)},scale=1.75]
      \foreach \t in {1,...,3}{
        \ngridLine{\t}{0}{\t}{1.75};
      }
      \mpsWire{0.25}{0}{3.75}{0};
      \mpsBvec{3.75}{0};
      \ngridLine{0.25}{1}{3.75}{1};
      \nME{3.75}{1};
      \sCircle{1}{1}{FcolU};
      \bCircle{2}{1}{FcolU};
      \sCircle{3}{1}{FcolU};
      \mpsA{1}{0}{colMPS};
      \mpsB{2}{0}{colMPS};
      \mpsA{3}{0}{colMPS};
    \end{tikzpicture}= {2}\,\,
    \begin{tikzpicture}
      [baseline={([yshift=-0.6ex]current bounding box.center)},scale=1.75]
      \foreach \t in {1,...,3}{
        \ngridLine{\t}{0}{\t}{1.25};
      }
      \mpsWire{0.25}{0}{3.75}{0};
      \mpsBvec{3.75}{0};
      \mpsB{3}{0}{colMPS};
      \mpsC{1}{2}{0}{colMPS};
      \ngridLine{1.5}{0.75}{0.25}{0.75};
      \sCircle{1}{0.75}{FcolU};
      \nclME{1.5}{0.75};
    \end{tikzpicture},\qquad
     \begin{tikzpicture}
      [baseline={([yshift=-0.6ex]current bounding box.center)},scale=1.75]
      \foreach \t in {1,...,2}{
        \ngridLine{\t}{0}{\t}{1.75};
      }
      \mpsWire{0.25}{0}{2.75}{0};
      \mpsBvec{2.75}{0};
      \mpsA{1}{0}{colMPS};
      \mpsB{2}{0}{colMPS};
      \ngridLine{0.25}{1}{2.75}{1};
      \sCircle{1}{1}{FcolU};
      \bCircle{2}{1}{FcolU};
      \nME{2.75}{1};
    \end{tikzpicture}= {2}\,\,
    \begin{tikzpicture}
      [baseline={([yshift=-0.6ex]current bounding box.center)},scale=1.75]
      \foreach \t in {1,...,2}{
        \ngridLine{\t}{0}{\t}{1.25};
      }
      \mpsWire{0.25}{0}{2.75}{0};
      \mpsBvec{2.75}{0};
      \mpsC{1}{2}{0}{colMPS};
      \ngridLine{1.5}{0.75}{0.25}{0.75};
      \sCircle{1}{0.75}{FcolU};
      \nclME{1.5}{0.75};
    \end{tikzpicture},   \\
      \label{eq:MagicRelations2}
    \begin{tikzpicture}[baseline={([yshift=-0.6ex]current bounding box.center)},scale=1.75]
      \foreach \t in {1,...,3}{
        \ngridLine{\t}{0}{\t}{1.75};
      }
      \mpsWire{0.25}{0}{3.75}{0};
      \mpsBvec{0.25}{0};
      \mpsB{1}{0}{colMPS};
      \mpsC{2}{3}{0}{colMPS};
      \ngridLine{0.25}{1}{2.5}{1};
      \nclME{2.5}{1};
      \sCircle{2}{1}{FcolU};
      \bCircle{1}{1}{FcolU};
      \nME{0.25}{1};
    \end{tikzpicture}=
    \begin{tikzpicture}[baseline={([yshift=-0.6ex]current bounding box.center)},scale=1.75]
      \foreach \t in {1,...,3}{
        \ngridLine{\t}{0}{\t}{0.5};
      }
      \mpsWire{0.25}{0}{3.75}{0};
      \mpsBvec{0.25}{0};
      \mpsA{1}{0}{colMPS};
      \mpsB{2}{0}{colMPS};
      \mpsA{3}{0}{colMPS};
    \end{tikzpicture},\quad  \begin{tikzpicture}
      [baseline={([yshift=-0.6ex]current bounding box.center)},scale=1.75]
      \foreach \t in {1,2}{
        \ngridLine{\t}{0}{\t}{1.25};
      }
      \mpsWire{0.25}{0}{2.75}{0};
      \mpsBvec{0.25}{0};
      \mpsC{1}{2}{0}{colMPS};
      \ngridLine{0.25}{0.75}{1.5}{0.75};
      \nME{0.25}{0.75};
      \sCircle{1}{0.75}{FcolU};
      \nclME{1.5}{0.75};
    \end{tikzpicture}=
    \begin{tikzpicture}
      [baseline={([yshift=-0.6ex]current bounding box.center)},scale=1.75]
      \foreach \t in {1,2}{
        \ngridLine{\t}{0}{\t}{0.5};
      }
      \mpsWire{0.25}{0}{2.75}{0};
      \mpsBvec{0.25}{0};
      \mpsB{1}{0}{colMPS};
      \mpsA{2}{0}{colMPS};
    \end{tikzpicture},\quad
    \begin{tikzpicture}
      [baseline={([yshift=-0.6ex]current bounding box.center)},scale=1.75]
      \foreach \t in {1,...,4}{
        \ngridLine{\t}{0}{\t}{1.75};
      }
      \mpsWire{0.25}{0}{4.75}{0};
      \mpsA{1}{0}{colMPS};
      \mpsB{2}{0}{colMPS};
      \mpsC{3}{4}{0}{colMPS};
      \ngridLine{0.25}{1}{3.5}{1};
      \sCircle{1}{1}{FcolU};
      \bCircle{2}{1}{FcolU};
      \sCircle{3}{1}{FcolU};
      \nclME{3.5}{1};
    \end{tikzpicture}=
    \begin{tikzpicture}
      [baseline={([yshift=-0.6ex]current bounding box.center)},scale=1.75]
      \foreach \t in {1,...,4}{
        \ngridLine{\t}{0}{\t}{1.25};
      }
      \mpsWire{0.25}{0}{4.75}{0};
      \mpsC{1}{2}{0}{colMPS};
      \mpsB{3}{0}{colMPS};
      \mpsA{4}{0}{colMPS};
      \ngridLine{0.25}{0.75}{1.5}{0.75};
      \sCircle{1}{0.75}{FcolU};
      \nclME{1.5}{0.75};
    \end{tikzpicture}.
  \end{align}
\end{subequations}
Here we introduced the ``classical'' maximum entropy state $\begin{tikzpicture}
  [baseline={([yshift=-0.6ex]current bounding box.center)},scale=1.75]
  \ngridLine{0}{0}{0.5}{0};
  \nclME{0.5}{0};
  \end{tikzpicture}=\begin{bmatrix}1&1&1&1\end{bmatrix}$ and the additional two-site tensor
\be
\begin{tikzpicture} [baseline={([yshift=-0.6ex]current bounding box.center)},scale=1.75]
  \mpsWire{-1}{4}{1}{4};
  \ngridLine{0.5}{4}{0.5}{4.75};
  \ngridLine{-0.5}{4}{-0.5}{4.75};
  \mpsC{-0.5}{0.5}{4}{colMPS};
  \node at ({5.5*\dx},{0.5*\dx}) {\scalebox{0.75}{$s_1\,r_1$}};
  \node at ({5.5*\dx},{-0.5*\dx}) {\scalebox{0.75}{$s_2\,r_2$}};
\end{tikzpicture}= C_{s_1 r_1 s_2 r_2}.
\ee 
In terms of these local tensors, the fixed point condition for the left eigenvector has a simple diagrammatic formulation. For example for $t=2$ we have 
\be
  \label{eq:fixedPointLinf}
  \!\!\!\bra{L_{\infty}}\tilde{\mathbb{W}}_{\infty}=
  \frac{1}{4}
\begin{tikzpicture}
  [baseline={([yshift=-1.6ex]current bounding box.center)},scale=1.7]
  \foreach \t in {1,...,4}{
    \ngridLine{\t}{0}{\t}{2.6};
  }
  \mpsWire{0.25}{0}{4.75}{0};
  \mpsBvec{0.25}{0};
  \mpsBvec{4.75}{0};
  \mpsB{1}{0}{colMPS};
  \mpsA{2}{0}{colMPS};
  \mpsB{3}{0}{colMPS};
  \mpsA{4}{0}{colMPS};
  \ngridLine{0.25}{1}{4.75}{1};
  \ngridLine{0.25}{2}{4.75}{2};
  \bCircle{1}{1}{FcolU};
  \sCircle{2}{1}{FcolU};
  \bCircle{3}{1}{FcolU};
  \sCircle{4}{1}{FcolU};
  \sCircle{1}{2}{FcolU};
  \bCircle{2}{2}{FcolU};
  \sCircle{3}{2}{FcolU};
  \bCircle{4}{2}{FcolU};
  \nME{0.25}{1};
  \nME{0.25}{2};
  \nME{4.75}{1};
  \nME{4.75}{2};
\end{tikzpicture}=
  \frac{1}{2}
\begin{tikzpicture}
  [baseline={([yshift=-1.6ex]current bounding box.center)},scale=1.7]
  \foreach \t in {1,...,4}{
    \ngridLine{\t}{0}{\t}{1.6};
  }
  \mpsWire{0.25}{0}{4.75}{0};
  \mpsBvec{0.25}{0};
  \mpsBvec{4.75}{0};
  \mpsA{1}{0}{colMPS};
  \mpsB{2}{0}{colMPS};
  \mpsA{3}{0}{colMPS};
  \mpsB{4}{0}{colMPS};
  \ngridLine{0.25}{1}{4.75}{1};
  \sCircle{1}{1}{FcolU};
  \bCircle{2}{1}{FcolU};
  \sCircle{3}{1}{FcolU};
  \bCircle{4}{1}{FcolU};
  \nME{0.25}{1};
  \nME{4.75}{1};
\end{tikzpicture}=
\begin{tikzpicture}
  [baseline={([yshift=-1.6ex]current bounding box.center)},scale=1.7]
  \foreach \t in {1,...,4}{
    \ngridLine{\t}{0}{\t}{0.6};
  }
  \mpsWire{0.25}{0}{4.75}{0};
  \mpsBvec{0.25}{0};
  \mpsBvec{4.75}{0};
  \mpsB{1}{0}{colMPS};
  \mpsA{2}{0}{colMPS};
  \mpsB{3}{0}{colMPS};
  \mpsA{4}{0}{colMPS};
\end{tikzpicture}=\bra{L_{\infty}}.
\ee
  The prefactor~${1}/{4}$ in the diagrammatic expression of~$\tilde{\mathbb{W}}_{\infty}$ comes from the normalisation of the infinite temperature state. To prove the fixed-point condition~\eqref{eq:fixedPointLinf}, we first apply the first relation of~\eqref{eq:MagicRelations1}, which replaces the bottom-most tensor~$A$ with~$B$, and introduces the two-site tensor~$C$ in the second and third leg. The tensor~$C$ is then repeatedly moved up using the third relation in~\eqref{eq:MagicRelations2}, until it is absorbed at the top by applying the left-most relation in~\eqref{eq:MagicRelations2}. This gives the MPS with the exchanged roles of~$A$ and~$B$. The procedure is then repeated, by using the right-most relation in~\eqref{eq:MagicRelations1} and the second identity in~\eqref{eq:MagicRelations2}, to complete the proof of~\eqref{eq:fixedPointLinf}. The form of the right leading vector~$\ket{R_{\infty}}$ is analogous, with the only difference that the roles of $A$ and $B$ are exchanged and the diagram is reflected (flipped from left to right). 

Besides giving access to all infinite temperature correlations, the expressions of $\bra{L_{\infty}}$ and $\ket{R_{\infty}}$ provide a natural basis to infer the structure of the fixed points $\bra{L}$ and $\ket{R}$ corresponding to thermalizing initial states. Specifically, we search for fixed points taking the same form up to the boundary vector $\ket{b}$ (cf. \eqref{eq:boundaryvectors}). This is because, at large times after the quench, the action of  the system on its own parts has to be indistinguishable from an infinite-temperature reservoir. 

To complete the ansatz, we then just have to specify a boundary vector for $\bra{L}$. This is done as follows. First, given the two-site shift invariance of the problem, we consider initial MPSs with this symmetry. Second, we observe that Eqs.~\eqref{eq:local_identities} define a fixed point of~$\tilde{\mathbb{W}}$ (with the graphical representation given in Fig.~\ref{fig:TN}), provided that the following boundary identities are fulfilled
\be
\label{eq:boundary_identities}
\begin{tikzpicture}
  [baseline={([yshift=-0.6ex]current bounding box.center)},scale=1.75]
  \foreach \t in {1,...,3}{
    \ngridLine{\t}{0}{\t}{1.75};
  }
  \mpsWire{0.25}{0}{3.75}{0};
  \vmpsWire{3.75}{0}{3.75}{1.75};
  \mpsA{1}{0}{colMPS};
  \mpsB{2}{0}{colMPS};
  \mpsA{3}{0}{colMPS};
  \ngridLine{0.25}{1}{3.75}{1};
  \sCircle{1}{1}{FcolU};
  \bCircle{2}{1}{FcolU};
  \sCircle{3}{1}{FcolU};
  \mpsBvecW{3.75}{0}{colMPS};
  \vmpsV{3.75}{1}{colVMPS}
\end{tikzpicture}=
\begin{tikzpicture}
  [baseline={([yshift=-0.6ex]current bounding box.center)},scale=1.75]
  \foreach \t in {1,...,3}{
    \ngridLine{\t}{0}{\t}{1.25};
  }
  \mpsWire{0.25}{0}{3.75}{0};
  \vmpsWire{3.75}{0}{3.75}{1.25};
  \mpsC{1}{2}{0}{colMPS};
  \mpsB{3}{0}{colMPS};
  \ngridLine{0.25}{0.75}{1.5}{0.75};
  \sCircle{1}{0.75}{FcolU};
  \nclME{1.5}{0.75}{FcolU};
  \mpsBvecV{3.75}{0}{colMPS};
\end{tikzpicture},\qquad
\begin{tikzpicture}
  [baseline={([yshift=-0.6ex]current bounding box.center)},scale=1.75]
  \foreach \t in {1,...,2}{
    \ngridLine{\t}{0}{\t}{1.75};
  }
  \mpsWire{0.25}{0}{2.75}{0};
  \vmpsWire{2.75}{0}{2.75}{1.75};
  \mpsA{1}{0}{colMPS};
  \mpsB{2}{0}{colMPS};
  \ngridLine{0.25}{1}{2.75}{1};
  \sCircle{1}{1}{FcolU};
  \bCircle{2}{1}{FcolU};
  \mpsBvecV{2.75}{0}{colMPS};
  \vmpsW{2.75}{1}{colVMPS}
\end{tikzpicture}=
\begin{tikzpicture}
  [baseline={([yshift=-0.6ex]current bounding box.center)},scale=1.75]
  \foreach \t in {1,...,2}{
    \ngridLine{\t}{0}{\t}{1.25};
  }
  \mpsWire{0.25}{0}{2.75}{0};
  \vmpsWire{2.75}{0}{2.75}{1.25};
  \mpsC{1}{2}{0}{colMPS};
  \ngridLine{1.5}{0.75}{0.25}{0.75};
  \nclME{1.5}{0.75}{FcolU};
  \sCircle{1}{0.75}{FcolU};
  \mpsBvecW{2.75}{0}{colMPS};
\end{tikzpicture},
\ee
where $\begin{tikzpicture}
  [baseline={([yshift=-0.6ex]current bounding box.center)},scale=1.75]
  \mpsWire{0}{0}{0.5}{0};
  \vmpsWire{0.5}{0}{0.5}{0.5};
  \mpsBvecV{0.5}{0}{colMPS};
  \node at ({0.75*\dx},{-0.5*\dx})  {\scalebox{0.8}{$\nu$}};
\end{tikzpicture} = \ket{v_{\nu}}$, 
$\begin{tikzpicture}
  [baseline={([yshift=-0.6ex]current bounding box.center)},scale=1.75]
  \mpsWire{0}{0}{0.5}{0};
  \vmpsWire{0.5}{0}{0.5}{0.5};
  \mpsBvecW{0.5}{0}{colMPS};
  \node at ({0.75*\dx},{-0.5*\dx})  {\scalebox{0.8}{$\nu$}};
\end{tikzpicture}=\ket{w_{\nu}}$, 
$\nu\in\{1,\ldots,\bondD\}$, are tensors to be determined. This gives 
\be
  \label{eq:fixedPointL}
  \bra{L}\tilde{\mathbb{W}}=
\begin{tikzpicture}
  [baseline={([yshift=-1.6ex]current bounding box.center)},scale=1.7]
  \foreach \t in {1,...,4}{
    \ngridLine{\t}{0}{\t}{2.6};
  }
  \vmpsWire{4.75}{0}{4.75}{2.55};
  \mpsWire{0.25}{0}{4.75}{0};
  \mpsBvec{0.25}{0};
  \mpsBvecW{4.75}{0}{colMPS};
  \mpsB{1}{0}{colMPS};
  \mpsA{2}{0}{colMPS};
  \mpsB{3}{0}{colMPS};
  \mpsA{4}{0}{colMPS};
  \ngridLine{0.25}{1}{4.75}{1};
  \ngridLine{0.25}{2}{4.75}{2};
  \bCircle{1}{1}{FcolU};
  \sCircle{2}{1}{FcolU};
  \bCircle{3}{1}{FcolU};
  \sCircle{4}{1}{FcolU};
  \sCircle{1}{2}{FcolU};
  \bCircle{2}{2}{FcolU};
  \sCircle{3}{2}{FcolU};
  \bCircle{4}{2}{FcolU};
  \nME{0.25}{1};
  \nME{0.25}{2};
  \vmpsV{4.75}{1}{colVMPS}
  \vmpsW{4.75}{2}{colVMPS};
\end{tikzpicture}=
\begin{tikzpicture}
  [baseline={([yshift=-1.6ex]current bounding box.center)},scale=1.7]
  \foreach \t in {1,...,4}{
    \ngridLine{\t}{0}{\t}{1.6};
  }
  \vmpsWire{4.75}{0}{4.75}{1.55};
  \mpsWire{0.25}{0}{4.75}{0};
  \mpsBvec{0.25}{0};
   \mpsBvecV{4.75}{0}{colMPS};
  \mpsA{1}{0}{colMPS};
  \mpsB{2}{0}{colMPS};
  \mpsA{3}{0}{colMPS};
  \mpsB{4}{0}{colMPS};
  \ngridLine{0.25}{1}{4.75}{1};
  \sCircle{1}{1}{FcolU};
  \bCircle{2}{1}{FcolU};
  \sCircle{3}{1}{FcolU};
  \bCircle{4}{1}{FcolU};
  \nME{0.25}{1};
   \vmpsW{4.75}{1}{colVMPS};
\end{tikzpicture}=
\begin{tikzpicture}
  [baseline={([yshift=-1.6ex]current bounding box.center)},scale=1.7]
  \foreach \t in {1,...,4}{
    \ngridLine{\t}{0}{\t}{0.6};
  }
  \vmpsWire{4.75}{0}{4.75}{.6};
  \mpsWire{0.25}{0}{4.75}{0};
  \mpsBvec{0.25}{0};
   \mpsBvecW{4.75}{0}{colMPS};
  \mpsB{1}{0}{colMPS};
  \mpsA{2}{0}{colMPS};
  \mpsB{3}{0}{colMPS};
  \mpsA{4}{0}{colMPS};
\end{tikzpicture}=\bra{L}.
\ee
Eqs.~\eqref{eq:boundary_identities} should be seen as a consistency equation for the bulk tensors defining the initial MPS, and the boundary vectors $\ket{v_{\nu}}$, $\ket{w_{\nu}}$. A priori, it is not obvious that a solution exists, but we find that this is indeed the case. In particular, considering the simplest case of initial product states ($\bondD=1$), given by a pair of one-site states, $\ket{\Psi_0} = \otimes_{j=1}^{L} \left(\ket{\psi_1}\otimes\ket{\psi_2}\right)$, $
\begin{tikzpicture}
  [baseline={([yshift=-1ex]current bounding box.center)},scale=1.75]
  \vmpsWire{0.5}{-0.5}{0.5}{0.5};
  \mpsWire{0}{0}{0.5}{0};
  \vmpsV{0.5}{0}{colVMPS};
\end{tikzpicture}=\ket{\psi_1}\ket{\psi_1}^{\ast}$,
$\begin{tikzpicture}
  [baseline={([yshift=-1ex]current bounding box.center)},scale=1.75]
  \vmpsWire{0.5}{-0.5}{0.5}{0.5};
  \mpsWire{0}{0}{0.5}{0};
  \vmpsW{0.5}{0}{colVMPS};
\end{tikzpicture}=\ket{\psi_2}\ket{\psi_2}^{\ast}$, the solution of Eq.~\eqref{eq:boundary_identities} exists for the one-parameter family of states
\be
\label{eq:initial_state}
\ket{\psi_{1}}=
\frac{1}{\sqrt{2}}\left(\ket{0}+\mathrm{e}^{\mathrm{i}\varphi}\ket{1}\right),\qquad
\ket{\psi_2}=\ket{0}.
\ee
The tensors $\ket{v_1}$, $\ket{w_1}$ are then uniquely determined~\cite{Note2}. This is our second main result: we have written the fixed points corresponding to the initial states \eqref{eq:initial_state} as MPSs with bond dimension three. Note that the fact that $\bra{L}$ and $\ket{R}$ are not product states implies that the dynamics is not purely Markovian, in contrast to the case of dual-unitary circuits~\cite{bertini2019entanglement,piroli2020exact}. Accordingly, the evolution of a given subsystem does not only depend on its state, and the quench protocol displays typical features of interacting many-body quantum systems.

In order to illustrate the power of our analytic solution, let us consider $\lim_{L\to\infty}\langle \Psi(t)|\mathcal{O}_{1;2x}|\Psi(t)\rangle$, where $\mathcal{O}_{1;2x}$ is a local operator spanning $2x$ sites. Making use of \eqref{eq:td_contraction} and representing the r.h.s.\ diagrammatically (cf.~\eqref{eq:fixedPointL}), we find that the expectation value can be expressed in terms of a time-independent map (matrix) $\mathcal{C}_{2x}$
\be\label{eq:one_point}
\bra{\Psi(t)}\mathcal{O}_{1;2x}\ket{\Psi(t)}=
\bra{\Phi[\mathcal{O}_{1;2x}]} \mathcal C^t_{2x} \ket{\Phi_{2x}},
\ee
where the matrix $\mathcal{C}_{2x}$ and vectors  $\bra{\Phi[\mathcal{O}_{1;2x}]}$,
$\ket{\Phi_{2x}}$ are defined graphically as
\be
  \label{eq:channelC}
  \!\!\!\!\!\!\!\mathcal C_{2x}  =
  \begin{tikzpicture}
    [baseline={([yshift=-2.4ex]current bounding box.center)},scale=1.4]
    \mpsWire{0}{0}{3}{0};
    \ngridLine{0}{1}{3}{1};
    \ngridLine{1}{0}{1}{1.75};
    \ngridLine{2}{0}{2}{1.75};
    \mpsA{2}{0}{colMPS};
    \mpsB{1}{0}{colMPS};
    \bCircle{1}{1}{FcolU};
    \sCircle{2}{1}{FcolU};
    \node at ({2.5*\dx},{-1.5*\dx}) {$\cdots$};
    \ngridLine{1}{3.25}{1}{5};
    \ngridLine{2}{3.25}{2}{5};
    \ngridLine{0}{4}{3}{4};
    \sCircle{1}{4}{FcolU};
    \bCircle{2}{4}{FcolU};
    \mpsWire{0}{5}{3}{5};
    \mpsB{2}{5}{colMPS};
    \mpsA{1}{5}{colMPS};
    \draw [decorate,decoration={brace,amplitude=5pt},xshift=0pt,yshift=0pt] 
    ({0.875*\dx},{0.125*\dx})--({4.125*\dx},{0.125*\dx}) node [midway,yshift=10pt]
    {$2x$};
  \end{tikzpicture},\,\,\,
  \begin{aligned}
\ket{\Phi_{2x}}\!&=\!\!
  \begin{tikzpicture}
    [baseline={([yshift=-2.4ex]current bounding box.center)},scale=1.5]
    \mpsWire{0}{0}{-0.75}{0};
    \mpsWire{0}{4}{-0.75}{4};
    \vmpsWire{0}{0}{0}{1.5};
    \vmpsWire{0}{2.5}{0}{4};
    \ngridLine{0}{1}{-0.75}{1};
    \ngridLine{0}{3}{-0.75}{3};
    \mpsBvecW{0}{0}{colMPS};
    \mpsBvecV{0}{4}{colMPS};
    \vmpsV{0}{1}{colVMPS};
    \vmpsW{0}{3}{colVMPS};
    \node at ({2*\dx},{0.5*\dx}) {$\cdots$};
    \node at ({2*\dx},{0*\dx}) {\scalebox{0.75}{$\cdots$}};
    \draw [decorate,decoration={brace,amplitude=5pt},xshift=0pt,yshift=0pt] 
    ({0.875*\dx},{0.875*\dx})--({3.125*\dx},{0.875*\dx}) node [midway,yshift=10pt]
    {$2x$};
  \end{tikzpicture}, \\
    \bra{\Phi[\mathcal{O}_{1;2x}]}\!&=\!
  \begin{tikzpicture}
    [baseline={([yshift=1.2ex]current bounding box.center)},scale=1.5]
    \mpsWire{0}{0}{0.75}{0};
    \mpsWire{0}{4}{0.75}{4};
    \mpsBvec{0}{0};
    \mpsBvec{0}{4};
    \ngridLine{0}{1}{0.75}{1};
    \ngridLine{0}{3}{0.75}{3};
    \extSquare{0}{1}{3}{colObs};
    \node at ({2*\dx},{-0.625*\dx}) {$\cdots$};
    \draw [decorate,decoration={brace,amplitude=5pt},xshift=0pt,yshift=0pt] 
    ({3.125*\dx},{-0.875*\dx})--({0.875*\dx},{-0.875*\dx}) node [midway,yshift=-10pt]
    {$2x$};
  \end{tikzpicture}.
  \end{aligned}
\ee
From Eqs.~\eqref{eq:one_point}--\eqref{eq:channelC}, we see that the time evolution of operators supported on a finite number of sites can be computed either analytically or in a numerically exact fashion, by diagonalising the matrix $\mathcal{C}_{2x}$. Indeed, since the computational cost of this operation does not depend on time, once this is done the dynamics can be followed for arbitrarily large times. We note, however, that since the matrix has dimensions ${9\cdot 2^{2x}}$, this procedure becomes exponentially costly as $x$ grows. Nevertheless, Eq.~\eqref{eq:one_point} can always be used to show that the late-time behaviour of any operator with finite support is exponentially decaying, and to compute exactly the corresponding characteristic time $\tau$. To see this, we use the identity
\be\label{eq:asymptotics}
 \!\!\!\!\!\!\! \begin{tikzpicture}
    [baseline={([yshift=-0.6ex]current bounding box.center)},scale=1.5]
    \vmpsWire{9}{0}{9}{5};
    \mpsWire{0.25}{0}{9}{0};
    \mpsWire{0.25}{5}{9}{5};
    \foreach \t in {1,...,8}{
      \ngridLine{\t}{0}{\t}{5};
    }
    \foreach \x in {1,...,4}{
      \ngridLine{0.25}{\x}{9}{\x};
    }
    \foreach \t in {1,3,...,8}{
      \mpsB{\t}{0}{colMPS};
      \mpsA{\t}{5}{colMPS};
      \bCircle{\t}{1}{FcolU};
      \sCircle{\t}{2}{FcolU};
      \bCircle{\t}{3}{FcolU};
      \sCircle{\t}{4}{FcolU};
    }
    \foreach \t in {2,4,...,8}{
      \mpsA{\t}{0}{colMPS};
      \mpsB{\t}{5}{colMPS};
      \sCircle{\t}{1}{FcolU};
      \bCircle{\t}{2}{FcolU};
      \sCircle{\t}{3}{FcolU};
      \bCircle{\t}{4}{FcolU};
    }
    \mpsBvecW{9}{0}{colMPS};
    \mpsBvecV{9}{5}{colMPS};
    \vmpsV{9}{1}{colVMPS};
    \vmpsW{9}{2}{colVMPS};
    \vmpsV{9}{3}{colVMPS};
    \vmpsW{9}{4}{colVMPS};
     \draw [decorate,decoration={brace,amplitude=5pt},xshift=0pt,yshift=0pt] 
    ({4*\dx},{-9.35*\dx})--({1*\dx},{-9.35*\dx}) node [midway,yshift=-10pt]
    {$2x$};
     \draw [decorate,decoration={brace,amplitude=5pt},xshift=0pt,yshift=0pt] 
    ({-0.5*\dx},{-8*\dx})--({-0.5*\dx},{-1*\dx}) node [midway,yshift=0pt,xshift=-12pt]
    {$2t$};
  \end{tikzpicture}=
  \begin{tikzpicture}
    [baseline={([yshift=-0.6ex]current bounding box.center)},scale=1.5]
    \vmpsWire{9}{0}{9}{5};
    \mpsWire{0.25}{0}{9}{0};
    \mpsWire{0.25}{5}{9}{5};
    \foreach \t in {1,...,8}{
      \ngridLine{\t}{0}{\t}{5};
    }
    \ngridLine{0.25}{1}{2.5}{1};
    \ngridLine{0.25}{2}{5.5}{2};
    \ngridLine{0.25}{3}{4.5}{3};
    \ngridLine{0.25}{4}{1.5}{4};

    \bCircle{1}{3}{FcolU};
    \bCircle{3}{3}{FcolU};
    \sCircle{1}{2}{FcolU};
    \sCircle{3}{2}{FcolU};
    \sCircle{5}{2}{FcolU};
    \bCircle{1}{1}{FcolU};
    \sCircle{1}{4}{FcolU};

    \sCircle{2}{3}{FcolU};
    \sCircle{4}{3}{FcolU};
    \bCircle{2}{2}{FcolU};
    \bCircle{4}{2}{FcolU};
    \sCircle{2}{1}{FcolU};

    \mpsB{1}{0}{colMPS};
    \mpsC{2}{3}{0}{colMPS};
    \mpsB{4}{0}{colMPS};
    \mpsC{5}{6}{0}{colMPS};
    \mpsB{7}{0}{colMPS};
    \mpsA{8}{0}{colMPS};

    \mpsC{1}{2}{5}{colMPS};
    \mpsB{3}{5}{colMPS};
    \mpsC{4}{5}{5}{colMPS};
    \mpsB{6}{5}{colMPS};
    \mpsA{7}{5}{colMPS};
    \mpsB{8}{5}{colMPS};

    \nclME{2.5}{1};
    \nclME{5.5}{2};
    \nclME{1.5}{4};
    \nclME{4.5}{3};
    \mpsBvecW{9}{0}{colMPS};
    \mpsBvecV{9}{5}{colMPS};
      \draw [decorate,decoration={brace,amplitude=5pt},xshift=0pt,yshift=0pt] 
    ({4*\dx},{-9.35*\dx})--({1*\dx},{-9.35*\dx}) node [midway,yshift=-10pt]
    {$2x$};
     \draw [decorate,decoration={brace,amplitude=3pt},xshift=0pt,yshift=0pt] 
    ({5.5*\dx},{-7*\dx})--({5.5*\dx},{-8*\dx}) node [midway,yshift=0pt,xshift=15pt]
    {$2t\!\!-\!\!3x\!\!$};
  \end{tikzpicture}
\ee
which can be proven graphically using Eqs.~\eqref{eq:local_identities} and~\eqref{eq:boundary_identities}. Eq.~\eqref{eq:asymptotics} implies that for any fixed $x$ the asymptotics of the expectation value is governed by the matrix $\mathcal C_0$ (cf. \eqref{eq:channelC}), independent of $x$. It is easy to show~\cite{Note2} that the spectrum of $\mathcal{C}_0$ is $\{1, \lambda,\lambda^*,0\}$, with $\lambda=(-3 + i \sqrt{7})/8$, while the only eigenvector associated with eigenvalue one is $\ket{b}\otimes\ket{b}$. This is the state that one would find at the bottom of the diagram on the l.h.s. of \eqref{eq:asymptotics} when computing the expectation value of $\mathcal{O}_{1;2x}$ in the infinite temperature state. This means that the local observable approaches exponentially its infinite-temperature stationary value
\be
\bra{\Psi(t)}\mathcal{O}_{1:2x}\ket{\Psi(t)}-{\rm tr}[\mathcal{O}_{1;2x} \rho_{\infty, 2x}]\sim \mathrm{e}^{-t/\tau}
\ee
where $\rho_{\infty,x}=\1/2^{x}$ is the infinite temperature state on $x$ sites and $\tau^{-1}=-2\log|\lambda|=2\log 2$. Thus, as anticipated before, 
 all local observables relax exponentially. This fact reflects the simple structure of our fixed points, and it is in contrast with the power-law relaxation displayed by some observables in more general integrable models~\cite{de2015relaxation}.

Eq.~\eqref{eq:one_point} can also be used to study (equal-time) two-point functions of generic local observables $a_x$ and $b_y$ (in this case one has to choose $\mathcal{O}_{x;y}=a_{x}b_{y}$) and explicitly identify a \emph{maximal} and a \emph{minimal} velocity of correlation spreading. Indeed, the brickwork structure imposes a maximal speed ${v_{\rm max}=2}$ (in our units), therefore for $t<x/2$ connected correlations are strictly zero. Additionally, Eq.~\eqref{eq:asymptotics} implies that, for $t>3x/2$, correlations are exponentially suppressed. Namely, there exist a minimal speed ${v_{\rm min}=2/3}$.  
Note that, accordingly, $v_{\rm min}$ and $v_{\rm max}$ are respectively the maximal and minimal value that the dressed velocity of quasiparticle excitations can take upon sampling all possible reference states~\cite{Friedman2019Integrable}.

\begin{figure}
  \includegraphics[scale=0.42]{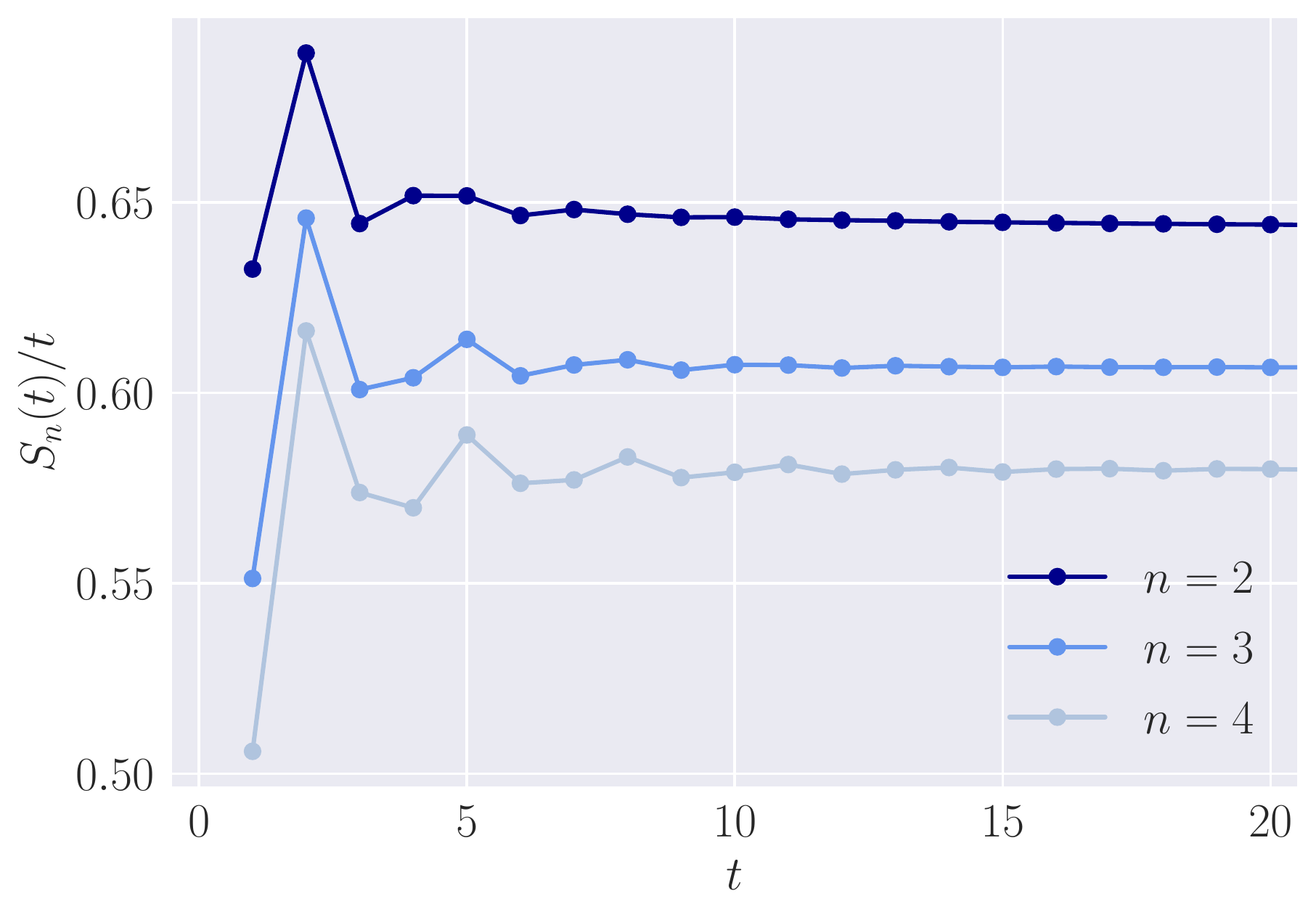}
  \caption{Growth of R\'enyi entropies, for $n=2,3,4$ and $\varphi=0$. The values of $S_n(t)$ are rescaled with time.}
  \label{fig:renyi}
\end{figure}

As a last physical application, we show that our solution allows us to compute explicitly the full time evolution  of the R\'enyi entropies after the quench. This is especially relevant in light of their accessibility in recent state-of-the-art quantum simulation experiments \cite{Greiner2015,Greiner2016,Linke2018,Greiner2019,Elben2020mixed}. By repeating the reasoning that led to Eq.~\eqref{eq:td_contraction} in the case of $2n$ copies (or ``replicas'') of the time-evolution operator ($n$ copies of $\mathbb{U}^t$ and $n$ copies of $(\mathbb{U}^\dag)^t$)~\cite{Note2} we find that $\bra{L}$ and $\ket{R}$ give access to the full time evolution of the entanglement entropies of a semi-infinite interval in the thermodynamic limit~\footnote{Here, for simplicity, we consider open boundary conditions.}. This is formalised by 
\be
S_{n}(t) = \frac{ \log[{\rm tr}[\rho^n_H(t)]]}{1-n} = \frac{\log[\bra{L}^{\otimes n} \mathcal S_{2n} (\ket{R}^*)^{\otimes n}]}{1-n} , 
\label{eq:entropy}
\ee
where $\rho_H(t)$ is the density matrix reduced to one of the halves, $ \mathcal S_{2n}$ is the operator performing a periodic shift by one copy in the $2n$-replica space~\cite{Note2}, and $(\cdot)^*$ denotes complex conjugation. 

Eq.~\eqref{eq:entropy} can again be written as the matrix element of the $t$-th power of a finite dimensional matrix. In particular, in this case the matrix is $3^n\times 3^n$ and we denote it by $\mathcal T_n$~\cite{Note2}. This means that for small $n$ we can compute $S_n(t)$ for arbitrarily large times: see, e.g., Fig.~\ref{fig:renyi}, where we report numerical data for $n=2,3,4$. As expected for interacting integrable systems~\cite{alba2017entanglement,alba2018entanglement,alba2017quench,alba2017renyi,mestyan2018renyi},  after an initial transient, we see a clear ballistic growth with asymptotic rate given by $
s_n\equiv \lim_{t \to \infty} S_{n}(t)/t= (\log \bar \lambda_n)/(1-n)$, where $\bar \lambda_n$ denotes the leading eigenvalue of $\mathcal T_n$. Remarkably, $\bar{\lambda}_n$ can be determined exactly for arbitrary $n$: we find that $\bar \lambda_n$ is the only real solution of  the equation
\be
4^n x^3-(x+1)^2=0.
\ee
    The proof of this statement is non-trivial, and will be reported elsewhere~\footnote{K. Klobas, and B. Bertini, in preparation.}. To the best of our knowledge, this result provides the first exact predictions for the asymptotic growth of R\'enyi entropies in an interacting (non-dual-unitary) system. Furthermore, it also allows us to perform an analytic continuation and obtain predictions for arbitrary (non-integer) values of $n$. In particular, taking the limit $n\to 1$, we obtain the asymptotic growth rate for the von Neumann entanglement entropy $s_1= \log 2$. This result is particularly significant, since it represents  the first exact confirmation of the quasiparticle picture for the entanglement spreading in the presence of interactions~\cite{alba2017entanglement, Note5}.\footnotetext[5]{In Rule 54 quasiparticles have no dispersion, the infinite-temperature dressed velocity is $1$~\cite{gopalakrishnan2018hydrodynamics} and the Yang-Yang entropy of the infinite temperature state is $\log 2$. Replacing these definitions in the prediction of Ref.~\cite{alba2017entanglement} gives us $s_1=\log 2$.} Namely, it proves that the growth of von Neumann entanglement entropy can be understood in terms of pairs of correlated quasiparticles created (at each point) by the quench and moving with opposite (dressed) velocities.

In summary, in this Letter we presented an exact description of the finite-time non-equilibrium dynamics generated by a quantum quench in an interacting integrable model: the quantum cellular automaton Rule 54. The fundamental ingredient for our derivations are a set of tensor network identities, the zipping conditions~\eqref{eq:boundaryvectors}, that enabled us to characterise exactly the action of the system on its own parts. It would be interesting to understand whether this approach can be extended to families of initial states relaxing to non-trivial GGEs, rather than to the infinite temperature state, which would enable the investigation of richer relaxation dynamics and more general dressing effects. For instance, this would provide new exact results for the spreading of entanglement, allowing us to perform a more extensive test of the quasiparticle picture for the dynamics of the von Neumann entropy and to derive analogous formulae for general R\'enyi entropies. Moreover, this would also enable an exact description of inhomogeneous quenches, paving the way for an ab initio derivation of Generalized Hydrodynamics~\cite{castroalvaredo2016emergent, bertini2016transport}. Another fascinating direction is to look for additional solutions of the zipping conditions. These would correspond to new exactly solvable (and possibly non-integrable) models for the discrete unitary dynamics beyond the dual-unitary case.

{\em Acknowledgments:}
We thank Toma{\v z} Prosen and Pasquale Calabrese for valuable comments on the manuscript.  LP acknowledges interesting discussions on related topics with Ignacio Cirac, Dmitry Abanin, Alessio  Lerose, and Michael Sonner. This work was supported by EPSRC under grant EP/S020527/1 (KK), the Royal Society through the University Research Fellowship No.\ 201101 (BB) and the Alexander von Humboldt foundation (LP).

\bibliography{./bibliography}

\onecolumngrid
\appendix
\newpage


\begin{center}
{\large{\bf Supplemental Material for\\
 Exact Thermalization Dynamics in the ``Rule $54$'' Quantum Cellular Automaton}}
\end{center}

Here we report some useful information complementing the main text. In particular
\begin{itemize}
\item[-] In Section~\ref{sec:tensornetwork} we derive the tensor-network representation of Rule 54;
\item[-] In Section~\ref{sec:spectrumTTM} we study the spectrum of the transfer matrix in space;
\item[-] In Section~\ref{sec:matrices} we provide the explicit form of matrices and boundary vectors defining the MPS representation of the fixed points
\item[-] In Section~\ref{sec:matC0} we report the matrix elements of~$\mathcal{C}_0$.
\item[-] In Section~\ref{sec:Renyi} we provide additional detail on the calculation of R\'enyi entropies. 
\end{itemize}

\section{Tensor network representation of Rule 54}
\label{sec:tensornetwork}

Time evolution in the model is obtained by consecutive application of the three-site local unitary gate $U$ given by~\eqref{eq:gateU}. Defining the following two tensors,
\be 
\begin{tikzpicture}
  [baseline={([yshift=-0.6ex]current bounding box.center)},scale=1.75]
  \gridLine{1}{0}{-1}{0};
  \gridLine{0}{1}{0}{-1};
  \bCircle{0}{0}{IcolU}
  \node at ({-1.3*\dx},{0}) {$s_1$};
  \node at (0,{-1.3*\dx}) {$s_2$};
  \node at ({1.3*\dx},{0}) {$\ s_3$};
  \node at (0,{1.3*\dx}) {$s_4$};
\end{tikzpicture}=
  \delta_{\chi(s_1,s_2,s_3),s_4}
  ,\qquad
  \begin{tikzpicture}
    [baseline={([yshift=-0.6ex]current bounding box.center)},scale=1.75]
    \def\sqrtThree{1.73205}
    \node at ({-\dx*1.3},0) {$s_1$};
    \node at ({-\dx/2.*1.3},{-\dx*\sqrtThree/2.*1.3}) {$s_2$};
    \node at ({\dx/2.*1.3},{-\dx*\sqrtThree/2.*1.3}) {$s_3$};
    \node at (0,{\dx*1.3}) {$s_k$};
    \node[label={[rotate=-75]{{$\cdots$}}},inner sep=0]
    (mid) at ({\dx/4.*2./3.},{\dx*2./3.*(1/2.-\sqrtThree/4)}) {};
    \gridLine{-0.75}{0}{0}{0}
    \gridLine{0}{-0.75}{0}{0}
    \gridLine{(\sqrtThree/2.*0.75)}{(-0.5*0.75)}{0}{0}
    \gridLine{(\sqrtThree/2.*0.75)}{(0.5*0.75)}{0}{0}
    \sCircle{0}{0}{IcolU};
  \end{tikzpicture} = \prod_{j=1}^{k-1} \delta_{s_{j},s_{j+1}},
\ee
one realizes that the three-site operator can be equivalently expressed as
\be
  U
  =\begin{tikzpicture}
    [baseline={([yshift=-0.6ex]current bounding box.center)},scale=1.75]
    \gridLine{-1}{1}{1}{1};
    \gridLine{-1}{2}{1}{2};
    \gridLine{-1}{3}{1}{3};
    \rcaGate{0}{1}{3}{IcolU}
  \end{tikzpicture}
  =\begin{tikzpicture}
[baseline={([yshift=-0.6ex]current bounding box.center)},scale=1.75]
    \gridLine{0}{1}{0}{3};
    \gridLine{-1}{1}{1}{1};
    \gridLine{-1}{2}{1}{2};
    \gridLine{-1}{3}{1}{3};
    \sCircle{0}{1}{IcolU};
    \bCircle{0}{2}{IcolU};
    \sCircle{0}{3}{IcolU};
  \end{tikzpicture}.
\ee 
The equivalence between the two sides in~Fig.~\ref{fig:circuit} then
follows from a simple identity satisfied by the second tensor,
\be
\begin{tikzpicture}
[baseline={([yshift=-0.6ex]current bounding box.center)},scale=1.75]
    \gridLine{-0.5}{0}{1.5}{0};
    \gridLine{1}{-1}{1}{0};
    \gridLine{0}{0}{0}{1};
    \sCircle{1}{0}{IcolU};
    \sCircle{0}{0}{IcolU};
  \end{tikzpicture}=
\begin{tikzpicture}
[baseline={([yshift=-0.6ex]current bounding box.center)},scale=1.75]
    \gridLine{-0.5}{0}{1.5}{0};
    \gridLine{0}{-1}{0}{0};
    \gridLine{1}{0}{1}{1};
    \sCircle{1}{0}{IcolU};
    \sCircle{0}{0}{IcolU};
  \end{tikzpicture}
  =
\begin{tikzpicture}
[baseline={([yshift=-0.6ex]current bounding box.center)},scale=1.75]
    \gridLine{-0.5}{0}{1.5}{0};
    \gridLine{0.5}{-1}{0.5}{1};
    \sCircle{0.5}{0}{IcolU};
  \end{tikzpicture}.
\ee

\section{Spectrum of the transverse transfer matrix}
\label{sec:spectrumTTM}

The unitarity of time-evolution implies
\be
1=\braket{\Psi(t)}{\Psi(t)}=\tr\left(\tilde{\mathbb{W}}^L\right)=\sum_{j} \lambda_j^L,
\ee
where the second equality follows directly from the definition of the transverse transfer matrix~\eqref{eq:transverse} and~$\lambda_j$ are eigenvalues of~$\tilde{\mathbb{W}}$. This equality holds for any~$L$, which means that~$\lambda_j\in\{0,1\}$ and both the geometric and algebraic multiplicity of the eigenvalue $1$ has to be~$1$.

\section{MPS representation of the leading eigenvectors}
\label{sec:matrices}

Here we report an explicit representation for the $3 \times 3$ matrices $A_{s r}$, $B_{s r}$, and $C_{s_1 r_1 s_2 r_2}$ fulfilling relations \eqref{eq:MagicRelations1}--\eqref{eq:MagicRelations2} and~\eqref{eq:boundary_identities}. The non identically vanishing matrices in the set are given by 
\be
\begin{aligned}
  A_{0 0}= \frac{1}{2}
  &\begin{bmatrix}
    1 & 1  & -1 \\
    1 & 1  &  1 \\
    1 & -1 & -1 
  \end{bmatrix}, &
  A_{0 1}=A_{1 0} = \frac{1}{2} 
  &\begin{bmatrix}
    0 & 1 & -1 \\
    1 & 0 & 0 \\
    1 & 0 & 0 
  \end{bmatrix}, &
  A_{1 1} =
  &\begin{bmatrix}
    0 & 1 & 0 \\
    1 & 0 & 0 \\
    0 & 0 & 0 
  \end{bmatrix},\\
  B_{0 0}=
  &\begin{bmatrix}
    1&0&0\\
    0&0&0\\
    0&0&0
  \end{bmatrix}, &
  B_{1 1}=
  &\begin{bmatrix}
    0&0&0\\
    0&1&0\\
    0&0&1
  \end{bmatrix},\\
  C_{0001}= C_{0010}=  \frac{1}{4}
  &\begin{bmatrix}
    0 & 1  & -1\\
    0 & 1  & -1\\
    0 & -1 & 1
  \end{bmatrix},&
  C_{0101}=C_{1010}= \frac{1}{4}
  &\begin{bmatrix}
    0 & 0& 0\\
    1 & 0& 0\\
    1 & 0& 0
  \end{bmatrix},&
  C_{0110}=C_{1001}= \frac{1}{4}
  &\begin{bmatrix}
    0 & 0 & 0\\
    0 & 1 & 1\\
    0 & 1 & 1
  \end{bmatrix},\\
  C_{1101}=C_{1110}=  \frac{1}{2}
  &\begin{bmatrix}
    1 & 0 & 0\\
    0 & 0 & 0\\
    0 & 0 & 0
  \end{bmatrix}, &
  C_{0 00 0} =
  \frac{1}{4}
  &\begin{bmatrix}
    1&1&-1 \\
    1&1&-1\\
    1&-1&1
  \end{bmatrix},&
  C_{0 01 1} =\frac{1}{2} 
  &\begin{bmatrix}
    0 & 1 & 0\\
    0 & 1 & 0\\
    0 & 0 & 1
  \end{bmatrix},\\
  C_{1100} = \frac{1}{2}
  &\begin{bmatrix}
    1 & 0 & 0\\
    0 & 1 & 1\\
    0 & 0 & 0
  \end{bmatrix},&
  C_{1111}=  \frac{1}{2}
  &\begin{bmatrix}
    1&0&0\\
    1&0&0\\
    0&0&0
  \end{bmatrix},
\end{aligned}
\ee
while the appropriate boundary vectors~$\ket{b}$,~$\ket{v_1}$, $\ket{w_1}$
are
\be
  \ket{b}=\frac{1}{\sqrt{2}}\begin{bmatrix}1\\1\\0\end{bmatrix},\qquad
  \ket{v_1}=\frac{1}{\sqrt{2}}\begin{bmatrix}1\\1\\-1\end{bmatrix},\qquad
    \ket{w_1}=\sqrt{2}\begin{bmatrix}1\\0\\0\end{bmatrix}.
\ee
Note that this choice of magnitude of boundary vectors implies the normalisation of leading eigenvectors of the transverse transfer matrix
\be
\braket{L_{\infty}}{R_{\infty}}=\braket{b}{b}^2=1,\qquad
\braket{L}{R}=\braket{b}{v}\braket{b}{w}=1.
\ee
The latter equations follow from Eq.~\eqref{eq:channelC} and $\bra{b}\otimes \bra{b} \mathcal{C}_0=\bra{b}\otimes \bra{b}$. 

\section{Spectrum of the matrix~$\mathcal{C}_0$}\label{sec:matC0}
By definition~\eqref{eq:channelC}, the matrix $\mathcal{C}_0$ can be expressed as
\be
\mathcal{C}_0=
\begin{tikzpicture}[baseline={([yshift=-0.6ex]current bounding box.center)},scale=1.75]
    \ngridLine{0}{0}{0}{1}
    \ngridLine{1}{0}{1}{1}
    \mpsWire{-0.75}{0}{1.75}{0};
    \mpsWire{-0.75}{1}{1.75}{1};
    \mpsB{0}{0}{colMPS}
    \mpsA{1}{0}{colMPS}
    \mpsA{0}{1}{colMPS}
    \mpsB{1}{1}{colMPS}
\end{tikzpicture}
=\frac{1}{4}
  \begin{bmatrix}
      1 & 0 & 0 & 1 &  2 & -2 & -1 & 0 & 0 \\
      1 & 0 & 0 & 1 &  2 &  2 & -1 & 0 & 0 \\
      1 & 0 & 0 & 1 & -2 & -2 & -1 & 0 & 0 \\
      0 & 4 & 0 & 0 &  0 &  0 &  0 & 0 & 0 \\
      2 & 0 & 0 & 2 &  0 &  0 &  2 & 0 & 0 \\
      0 & 0 & 0 & 0 &  0 &  0 &  0 & 0 & 0 \\
      0 & 0 & 0 & 0 &  0 &  0 &  0 & 0 & 0 \\
      2 & 0 & 0 &-2 &  0 &  0 & -2 & 0 & 0 \\
      0 & 0 & 0 & 0 &  0 &  0 &  0 & 0 & 0 
  \end{bmatrix}.
\ee
It is easy to verify that it has $3$ non-zero eigenvalues with geometric and algebraic multiplicities equal to $1$. Explicitly
\be
  \mathrm{Sp}\left(\mathcal{C}_0\right) =
  \{1,-\frac{3}{8}+\mathrm{i}\frac{\sqrt{7}}{8},
  -\frac{3}{8}-\mathrm{i}\frac{\sqrt{7}}{8},0\}.
\ee

\section{Tensor network formulation of R\'enyi entanglement entropies}
\label{sec:Renyi}

We start by considering the density matrix at time $t$ (at time 0 the system has been initialised in an MPS fulfilling~\eqref{eq:boundary_identities}) reduced to one of the halves
\be \label{eq:rho_h}
\rho_H(t)=\frac{1}{\mathcal N_L}\quad
\begin{tikzpicture}
  [baseline={([yshift=-0.6ex]current bounding box.center)},scale=1.75]
  \newcommand\bCross[2]{
    \draw [thick,colLines] ({\dx*#2-0.5*\r},{-\dx*#1-0.5*\r}) --
    ({\dx*#2+0.5*\r},{-\dx*#1+0.5*\r});
    \draw [thick,colLines] ({\dx*#2-0.5*\r},{-\dx*#1+0.5*\r}) --
    ({\dx*#2+0.5*\r},{-\dx*#1-0.5*\r});
  }
  \IvmpsWire{0}{0.25}{0}{8.75};
  \IvmpsWire{0.75}{0.25}{0.75}{8.75};
  \bCross{0}{0.25};
  \bCross{0}{8.75};
  \bCross{0.75}{0.25};
  \bCross{0.75}{8.75};
  \foreach \t in {-1,...,-4}{
    \gridLine{\t}{0.25}{\t}{8.75};
    \bCross{\t}{0.25};
    \bCross{\t}{8.75};
  }
  \foreach \t in {1.75,...,4.75}{
    \gridLine{\t}{0.25}{\t}{8.75};
    \bCross{\t}{0.25};
    \bCross{\t}{8.75};
  }
  \foreach \x in {1,...,8}{
    \gridLine{0}{\x}{-4.75}{\x};
    \gridLine{0.75}{\x}{5.5}{\x};
  }
  \foreach \x in {1,3,...,8}{
    \vmpsW{0}{\x}{IcolVMPS}
    \vmpsW{0.75}{\x}{IcolVMPSc}
    \sCircle{-1}{\x}{IcolU}
    \bCircle{-2}{\x}{IcolU}
    \sCircle{-3}{\x}{IcolU}
    \bCircle{-4}{\x}{IcolU}
    \sCircle{1.75}{\x}{IcolUc}
    \bCircle{2.75}{\x}{IcolUc}
    \sCircle{3.75}{\x}{IcolUc}
    \bCircle{4.75}{\x}{IcolUc}
  }
  \foreach \x in {2,4,...,8}{
    \vmpsV{0}{\x}{IcolVMPS}
    \vmpsV{0.75}{\x}{IcolVMPSc}
    \bCircle{-1}{\x}{IcolU}
    \sCircle{-2}{\x}{IcolU}
    \bCircle{-3}{\x}{IcolU}
    \sCircle{-4}{\x}{IcolU}
    \bCircle{1.75}{\x}{IcolUc}
    \sCircle{2.75}{\x}{IcolUc}
    \bCircle{3.75}{\x}{IcolUc}
    \sCircle{4.75}{\x}{IcolUc}
  }
  \foreach \x in {1,...,4}{
    \gridLine{-4.75}{(-0.1-0.2*\x)}{5.5}{(-0.1-0.2*\x)};
    \draw [thin,colLines,rounded corners=2] ({(-0.1-0.2*\x)*\dx},{4.75*\dx}) --
    ({(-0.1-0.2*\x)*\dx},{(4.9+0.1*\x)*\dx}) --
    ({\x*\dx},{(4.9+0.1*\x)*\dx}) -- ({\x*\dx},{4.75*\dx});
    \draw [thin,colLines,rounded corners=2] ({(-0.1-0.2*\x)*\dx},{-5.5*\dx}) --
    ({(-0.1-0.2*\x)*\dx},{-(5.6+0.1*\x)*\dx}) --
    ({\x*\dx},{-(5.6+0.1*\x)*\dx}) -- ({\x*\dx},{-5.5*\dx});
  }
\end{tikzpicture}\,.
\ee
Here we took \emph{open} boundary conditions and introduced the constant $\mathcal N_L$ to ensure ${{\rm tr}[\rho_H(t)]=1}$. Note that $\lim_{L\to\infty}\mathcal N_L = (\bra{B_L}\otimes\bra{B_L}^*)\ket{R}\bra{L}(\ket{B_R}\otimes\ket{B_R}^*)$ where $\bra{B_L}, \ket{B_R}\in \mathbb C^{2^{2t}}$ are the vectors in the $t$-direction encoding the (left and right) boundary conditions, the tensor product acts on different copies, and $(\cdot)^*$ denotes complex conjugation. Note that  $\bra{B_L}$ and $\ket{B_R}$ are defined in one copy of the time-sheet, while $\bra{L}, \ket{R} \in \mathbb C^{2^{4t}}$ live in the tensor product of two copies. 

Considering now the trace of, e.g., the third power of the reduced density matrix we have 
\be
{\rm tr}[\rho^3_H(t)]=\frac{1}{\mathcal N^3_L} \quad
\begin{tikzpicture}
  [baseline={([yshift=-0.6ex]current bounding box.center)},scale=1.75]
  \TECsheet{-2.5}{2.5};
  \TEsheet{-2}{2};
  \TECsheet{-1.5}{1.5};
  \TEsheet{-1}{1};
  \TECsheet{-0.5}{0.5};
  \TEsheet{0}{0};
  \bConnection{-2.5}{6}{-2.75}{6.25};
  \connection{-2}{2}{-2.5}{2.5};
  \connection{-1.5}{5}{-2}{5.5};
  \connection{-1}{1}{-1.5}{1.5};
  \connection{-0.5}{4}{-1}{4.5};
  \connection{0}{0}{-0.5}{0.5};
  \fConnection{0.4}{3.1}{0}{3.5};
\end{tikzpicture}\,,
\ee
where green and red areas represent the TNs associated with $U$ and $U^\dagger$ respectively, cf. Eq.~\eqref{eq:rho_h}. Using the t-channel language of Eq.~\eqref{eq:transverse}, we see that this object can be written as 
\be
{\rm tr}[\rho^3_H(t)]=\frac{1}{\mathcal N^3_L} (\bra{B_L}\otimes\bra{B_L}^*)^{\otimes 3} (\tilde{\mathbb{W}}^{\otimes 3})^{L} \mathcal S_{6} ((\tilde{\mathbb{W}}^*)^{\otimes 3})^{L} (\ket{B_R}^*\otimes\ket{B_R}),
\ee
where $\mathcal{S}_6$ is the periodic shift by one in the space of the $6$ replicas. Namely it acts on the Hilbert space 
\be
\mathcal H_6=\bigl(\mathbb C^{2^{2t}}\bigr)^{\otimes 6}
\ee 
as follows 
\be
 \mathcal S_{6}  \ket{i_1}\otimes\ket{i_2}\otimes\ket{i_3}\otimes\ket{i_4}\otimes\ket{i_5}\otimes\ket{i_6} = \ket{i_2}\otimes\ket{i_3}\otimes\ket{i_4}\otimes\ket{i_5}\otimes\ket{i_6}\otimes\ket{i_1}\,,\qquad i_j=0,1,\ldots 2^{2t}-1\,,
\ee
where $\{\ket{i}\}_{i=0}^{2^{2t}-1}$ is a basis of $\mathbb C^{2^{2t}}$.

In the thermodynamic limit we can make the replacement  
\be
\tilde{\mathbb{W}} \mapsto \ket{R}\!\bra{L}\,,\qquad \tilde{\mathbb{W}}^* \mapsto \ket{R}^*\!\bra{L}^*\,,\qquad \mathcal N_L\mapsto   (\bra{B_L}\otimes\bra{B_L}^*)\ket{R}\bra{L}(\ket{B_R}\otimes\ket{B_R}^*),
\ee
and use 
\be
(\bra{B_L}\otimes\bra{B_L}^*)\ket{R} = (\bra{B_L}^*\otimes\bra{B_L})\ket{R}^*\,,\qquad\bra{L}(\ket{B_R}\otimes\ket{B_R}^*)= \bra{L}^*(\ket{B_R}^*\otimes\ket{B_R}),
\ee
to find 
\be
\lim_{L\to\infty} {\rm tr}[\rho^3_H(t)] = \bra{L}^{\otimes 3} \mathcal S_{6} (\ket{R}^*)^{\otimes 3}\,.
\label{eq:thermotracered}
\ee
An analogous reasoning considering a generic $n$-th power leads to \eqref{eq:entropy}. Note that for periodic boundary conditions the r.h.s.\ of \eqref{eq:thermotracered} is replaced by its square.  

Finally, we note that \eqref{eq:thermotracered} can be represented in terms of the matrix element of the $t$-th power of a time-independent matrix. Indeed, ``unfolding" the tensors \eqref{eq:boundaryvectors}, i.e.
\be
  \begin{tikzpicture}
    [baseline={([yshift=-0.6ex]current bounding box.center)},scale=1.75]
    \mpsWire{-0.75}{0}{0.75}{0};
    \ngridLine{0}{0}{0}{0.75};
    \mpsA{0}{0}{colMPS}
    \node at ({1.5*\dx},0) {$s_1s_2$};
  \end{tikzpicture} \rightarrow
  \begin{tikzpicture}
    [baseline={([yshift=-0.6ex]current bounding box.center)},scale=1.75]
    \mpsWire{-0.75}{0}{0.75}{0};
    \gridLine{0}{-0.75}{0}{0.75};
    \mpsA{0}{0}{colMPS}
    \node at ({-1.125*\dx},0) {$s_1$};
    \node at ({1.125*\dx},0) {$s_2$};
  \end{tikzpicture},\qquad
  \begin{tikzpicture}
    [baseline={([yshift=-0.6ex]current bounding box.center)},scale=1.75]
    \mpsWire{-0.75}{0}{0.75}{0};
    \ngridLine{0}{0}{0}{0.75};
    \mpsB{0}{0}{colMPS}
    \node at ({1.5*\dx},0) {$s_1s_2$};
  \end{tikzpicture} \rightarrow
  \begin{tikzpicture}
    [baseline={([yshift=-0.6ex]current bounding box.center)},scale=1.75]
    \mpsWire{-0.75}{0}{0.75}{0};
    \gridLine{0}{-0.75}{0}{0.75};
    \mpsB{0}{0}{colMPS}
    \node at ({-1.125*\dx},0) {$s_1$};
    \node at ({1.125*\dx},0) {$s_2$};
  \end{tikzpicture},
\ee
we have 
\be
\bra{L}^{\otimes 3} \mathcal S_{6} (\ket{R}^*)^{\otimes 3} = \bra{\mathcal U_{3}} \mathcal T^t_{3} \ket{\mathcal D_{3}},
\ee
with  
\be
\mathcal{T}_n \equiv 
  \begin{tikzpicture}
    [baseline={([yshift=-2.6ex]current bounding box.center)},scale=1.75]
    \gridLine{0}{0.25}{0}{6.75};
    \gridLine{1}{0.25}{1}{6.75};
    \leftHook{0}{0.25};
    \rightHook{0}{6.75};
    \leftHook{1}{0.25};
    \rightHook{1}{6.75};
    \foreach \x in {1,3,5}{
      \mpsWire{-0.75}{\x}{1.75}{\x};
      \mpsA{1}{\x}{colMPS};
      \mpsB{0}{\x}{colMPS};
    }
    \foreach \x in {2,4,6}{
      \mpsWire{-0.75}{\x}{1.75}{\x};
      \mpsA{0}{\x}{colMPS};
      \mpsB{1}{\x}{colMPS};
    }
      \draw [decorate,decoration={brace,amplitude=5pt},xshift=0pt,yshift=0pt] 
    ({0.875*\dx},{0.8*\dx})--({6.125*\dx},{0.8*\dx}) node [midway,yshift=10pt]
    {$2n$};
  \end{tikzpicture},\qquad
 \ket{\mathcal D_n} \equiv \begin{tikzpicture}
    [baseline={([yshift=-2.6ex]current bounding box.center)},scale=1.75]
    \foreach \x in {0,...,5}{
    \mpsWire{0}{\x}{-0.75}{\x};
  }
    \vmpsWire{0}{0}{0}{5};
    \mpsBvecW{0}{0}{colMPS};
    \mpsBvecV{0}{1}{colMPS};
    \mpsBvecW{0}{2}{colMPS};
    \mpsBvecV{0}{3}{colMPS};
    \mpsBvecW{0}{4}{colMPS};
    \mpsBvecV{0}{5}{colMPS};
     \draw [decorate,decoration={brace,amplitude=5pt},xshift=0pt,yshift=0pt] 
    ({0.875*\dx-1*\dx},{0.8*\dx})--({5.125*\dx},{0.8*\dx}) node [midway,yshift=10pt]
    {$2n$};
  \end{tikzpicture},\qquad
 \bra{\mathcal U_n}  \equiv \begin{tikzpicture}
    [baseline={([yshift=1.2ex]current bounding box.center)},scale=1.75]
    \foreach \x in {0,...,5}{
      \mpsWire{0}{\x}{0.75}{\x};
      \mpsBvec{0}{\x};
    }
      \draw [decorate,decoration={brace,amplitude=5pt},xshift=0pt,yshift=0pt] 
    ({5.125*\dx},{-0.8*\dx}) -- ({0.875*\dx-1*\dx},{-0.8*\dx}) node [midway,yshift=-10pt]
    {$2n$};
  \end{tikzpicture}.
\ee

\end{document}